\documentclass[twocolumn,prl,superscriptaddress,aps,amsmath,floatfix]{revtex4-1}

\usepackage{graphics}
\usepackage{graphicx}
\usepackage{epstopdf}
\usepackage{dcolumn}
\usepackage{bm}
\usepackage{longtable}
\usepackage{epsfig}
\usepackage{times}
\usepackage{url}
\usepackage{color}

\begin{document}
\title{Time-Delayed Magnetic Control and Narrowing of X-Ray frequency Spectra in Two-Target Nuclear Forward Scattering}
\author{Po-Han   \surname{Lin}}
\affiliation{Department of Physics, National Central University, Taoyuan City 32001, Taiwan}

\author{Yen-Yu  \surname{Fu}}
\affiliation{Department of Physics, National Central University, Taoyuan City 32001, Taiwan}

\author{Wen-Te \surname{Liao}}
\email{wente.liao@g.ncu.edu.tw}
\affiliation{Department of Physics, National Central University, Taoyuan City 32001, Taiwan}
\date{\today}
\begin{abstract}
Controlling and narrowing x-ray frequency spectra in magnetically perturbed two-target nuclear forward scattering is theoretically studied. 
We show that different hard-x-ray  spectral redistributions can be achieved by single or multiple  switching of magnetic field in nuclear targets.
Our scheme can generate x-ray spectral lines with tenfold intensity enhancement  and spectral width narrower than four times the nuclear natural linewidth.
The present results pave the way towards a brighter and flexible x-ray source for  precision spectroscopy of nuclear resonances using modern synchrotron radiation.
\end{abstract}

\maketitle

Dynamic perturbations on low-lying nuclear energy levels lead to  modulations of  M\"ossbauer frequency spectrum \cite{Ikonen1988, Vagizov1990, Tittonen1992, Olga1999, Roehlsberger2012, Heeg2017, Ramien2018} and controls of time spectrum \cite{Helisto1982, Smirnov1984, Helisto1991, Shvydko1996, Smirnov1996a, Palffy2009, Liao2012a, Olga2014, Liao2014, Heeg2015a, Kong2016, Liao2017, Wang2018} in  x-ray quantum optics \cite{Adams2013}.
These  perturbations are mainly implemented by either mechanical vibrations \cite{Helisto1982, Smirnov1996a,  Shakhmuratov2013,Olga2014, Heeg2017} or magnetic switching \cite{Ikonen1988, Vagizov1990, Tittonen1992, Shvydko1996}, which are two  effective ways to  tame  photons in keV region coupling to nuclear transitions.
Coherent control of hard x rays allows for advances  in understanding of fundamental physics \cite{Buervenich2006, Roehlsberger2010, Liao2011,Liao2015} and information technology \cite{Palffy2009,  Liao2012a, Liao2014,  Olga2014,  Wang2018, Zhang2019}.
Moreover, spectral narrowing of x-ray synchrotron radiation pulses  in nuclear forward scattering (NFS) can be achieved by utilizing the fast and fine mechanical motion of a resonant target \cite{Heeg2017}.
This on one hand remarkably offers the opportunity of pushing nuclear x-ray spectroscopy toward higher resolution by transferring  off-resonant photons onto the tiny resonant fraction, but on the other hand needs the expense of sophisticated control over the target trajectory  \cite{Heeg2017}. 
Here, we theoretically demonstrate that, without any precise mechanical control, a magnetically perturbed two-target NFS system enhances x-ray spectral intensity several times the  overwhelming off-resonant background within a spectral width of 4 times the nuclear natural linewidth.
Furthermore, time-delayed magnetic switching  causes an alternation of intensity at different frequencies, and so offers a flexible selection of spectral lines. 
Our chronological magnetic switching provides any precision spectroscopy of ultranarrow nuclear resonances with brighter and controllable x-ray source.
%

\begin{figure}[b]
\includegraphics[width=0.43\textwidth]{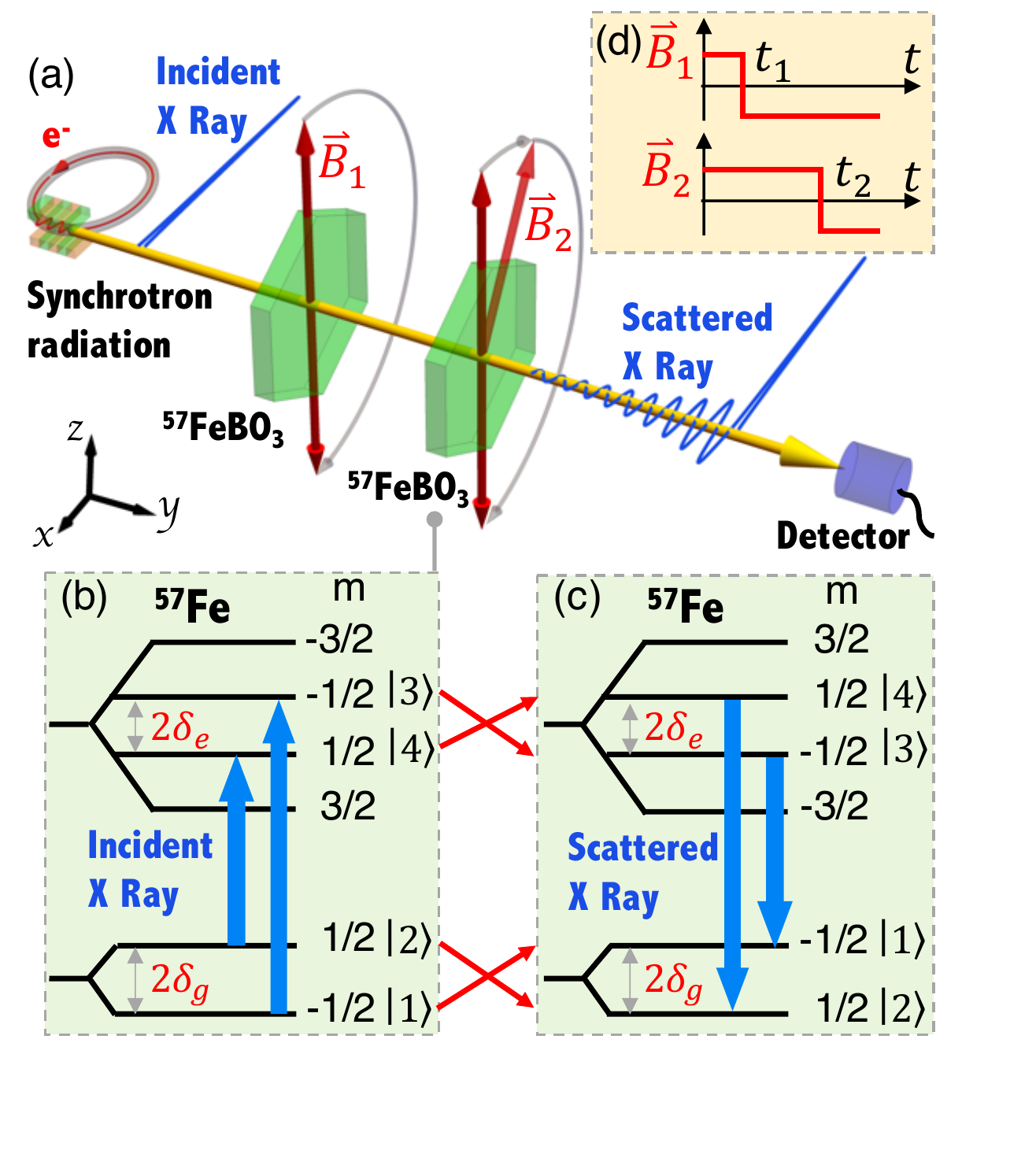}
\caption{\label{fig1}
(Color online) (a) A linearly polarized  x-ray synchrotron radiation of 14.4 keV (blue curves) impinges on two $^{57}$FeBO$_3$ crystals (green hexagon) applied with external magnetic field $\vec{B}_1$ and $\vec{B}_2$ (red arrows).
(b) $^{57}$Fe nuclear energy level with ground (excited) state Zeeman shift $\delta_g$ ($\delta_e$). 
The nuclear spin projection on z axis is denoted by $m$.
The x ray drives $\vert 1\rangle \rightarrow \vert 3\rangle $ and $\vert 2\rangle \rightarrow \vert 4\rangle $  transitions. By turning over the magnetic field $\vec{B}\rightarrow -\vec{B}$, the energy levels are shifted as illustrated in (c).
(d) $\vec{B}_1$ and $\vec{B}_2$ are inverted at $t=t_1$ and $t=t_2$, respectively, with a time delay $\tau_D=t_2-t_1$.
}
\end{figure}
Figure~\ref{fig1}(a) illustrates the two-target NFS system, and Fig.~\ref{fig1}(b) depicts the $^{57}$Fe nuclear-level scheme \cite{Shvydko1996, Liao2012a}. Two $^{57}$Fe isotopically enriched $^{57}$FeBO$_3$ crystals are impinged by a  linearly polarized x ray of 14.4-keV photon energy. The incident x ray drives $\Delta m=0$ transitions, namely, $\vert 1\rangle \rightarrow \vert 3\rangle $ and $\vert 2\rangle \rightarrow \vert 4\rangle $ transitions. Here $m$ is the nuclear spin projection along  the z axis.
$\delta_g$ and $\delta_e$ are the hyperfine splitting for ground states and excited states, respectively. 
When applying external magnetic fields $\vec{B}_1$ and $\vec{B}_2$ of few tens Gauss in the easy plane of a $^{57}$FeBO$_3$ crystal, the orientation of the crystal magnetization can be easily switched by inverting each magnetic field within few ns (the red upward arrows become the downward arrows) \cite{Shvydko1996}. 
When turning over the magnetic field the energy levels will exchange as demonstrated by Fig.~\ref{fig1}(c).  
As illuistrated in Fig.~\ref{fig1}(d),  in a two-target NFS system one can  switch $\vec{B}_1$ at $t=t_1$ and $\vec{B}_2$ at $t=t_2$ with a time delay $\tau_D=t_2-t_1$.
We will show that such time-delayed magnetic control 
leads to  versatile manipulations of NFS  frequency spectrum.
The following optical-Bloch equation (OBE) describes our two-target NFS system  \cite{Liao2012a, Kong2014, Wang2018, Zhang2019}:
\begin{eqnarray}
 &  & \partial_{t}{\rho_{31}^{(j)}} =-[\frac{\Gamma}{2}+i\Delta M_j(t)]\rho_{31}^{(j)}+i\frac{a}{4}\Omega_j, \label{eq1}\\
 &  & \partial_{t}{\rho_{42}^{(j)}} =-[\frac{\Gamma}{2}-i\Delta M_j(t)]\rho_{42}^{(j)}+i\frac{a}{4}\Omega_j, \label{eq2}\\
 &  & \frac{1}{c}\partial_{t} \Omega_j + \partial_{y} \Omega_j =i\eta_j( \rho_{31}^{(j)}+\rho_{42}^{(j)})-\frac{k}{2i}(n^2-1)\Omega_j . \label{eq3}
\end{eqnarray}%
Here index $j\in \left\lbrace 1, 2\right\rbrace $ indicates the quantity for the $j$th target.
$\Omega_j$ is the x-ray Rabi frequency,  and 
$\rho_{31}^{(j)}$ and $\rho_{42}^{(j)}$ are the off-diagonal density matrix elements for the four-level $^{57}$Fe nuclear level scheme depicted in Fig.~\ref{fig1}(b).
$\Gamma=1/141$ GHz is the spontaneous decay rate of excited states $\vert{3}\rangle$ and $\vert{4}\rangle$. $\Delta = \delta_g+\delta_e$ represents the x-ray detuning due to hyperfine splitting. The dynamical magnetic switching is described by $M_j(t)=-\tanh\left( \frac{t-t_j}{0.25d}\right) $, where the external magnetic field $\vec{B}_j$ is inverted at $t=t_j$ within a duration of $d=2$ns.  
$c$ is the speed of light in vacuum, $\eta_j=2\Gamma\xi_j/\left( a L\right) $, and $\xi_j$ is the nuclear resonant thickness of the $j$th target. $a=\sqrt{2/3}$ is the Clebsch-Gordan coefficients, and $L$ is the sample thickness. $k$ is the x-ray wave number, and $n\approx 1+ i 9.13\times 10^{-8} $ is the x-ray refractive index of a $^{57}$FeBO$_3$ crystal contributed by electrons \cite{nndc, lbl, ipmt}. 
The initial and boundary  conditions are $\rho_{31}^{(j)}\left( y, 0\right) = \rho_{42}^{(j)}\left( y, 0\right) = \Omega_j \left( y, 0 \right) =0$, $\Omega_1 \left( 0, t\right) = \exp\left[ -\left( \frac{t-t_0}{\tau}  \right) ^2  \right] $, and $\Omega_2 \left( 0, t \right) =\Omega_1 \left( L, t\right) $, where ($t_0$, $\tau$) = ($0.67$ns, $0.1$ns).
The normalized output  frequency spectrum $S(\omega)$  is determined by
\begin{equation}\label{eq4}
S(\omega)=\frac{\lvert\int_{-\infty}^\infty\Omega_2(L,t)e^{i\omega t} dt\rvert^{2}}{\max\lvert\int_{-\infty}^\infty \Omega_1(0, t)e^{i\omega t} dt \rvert^{2}} .
\end{equation}
%

%

When an ultrashort x-ray pulse $\delta(t)$  illuminates an one-target or a two-target system, the  scattered field is respectively the real part of
\begin{eqnarray}
E_1(t) &=& \delta(t)-W_1(M_1,t) , \label{eq5}\\ 
E_2(t) &=& \delta(t)-W_1(M_1,t)-W_2(M_2,t)\nonumber \\
&+&\int_{0}^{t}W_2(M_2,t-t')W_1(M_1, t')dt'. \label{eq6}
\end{eqnarray}
Here $W_j( 1, t)=\frac{\xi_j}{\sqrt{\xi_j\Gamma t}}J_{1}\left( 2\sqrt{\xi_j\Gamma t}\right)  e^{-\frac{\Gamma}{2}t+i\Delta t}$ is  the single-target response function,  and $J_{1}$ is the Bessel function of first kind  for the unperturbed NFS.
The Bessel function depicts the multiple scattering, namely, dynamical beat (DB), whose frequency is proportional to $\xi_j$. 
$\cos\left( \Delta t\right)$ illustrates the oscillation from hyperfine splitting $\Delta$, i.e.,
quantum beat (QB)  \cite{Helisto1991, Shvydko1998, Shvydko1999N, Roehlsberger2004}.
In order to avoid the complicated hybrid beat \cite{Shvydko1998}, we focus on
either $\Delta/\Gamma > \xi_j$ or $\Delta/\Gamma < \xi_j$.
%
The interference of four scattering paths constitutes  Eq.~(\ref{eq6}) \cite{Roehlsberger2004, Smirnov2005, Liao2012a}. $\delta(t)$ represents that no scattering occurs. $-W_j(M_j,t)$ depicts that only the $j$th target scatters x rays. The last convolution term describes that x rays are chronologically scattered by the first target and then by the second.
%
%
%
%

\begin{figure}[b]
\includegraphics[width=0.43\textwidth]{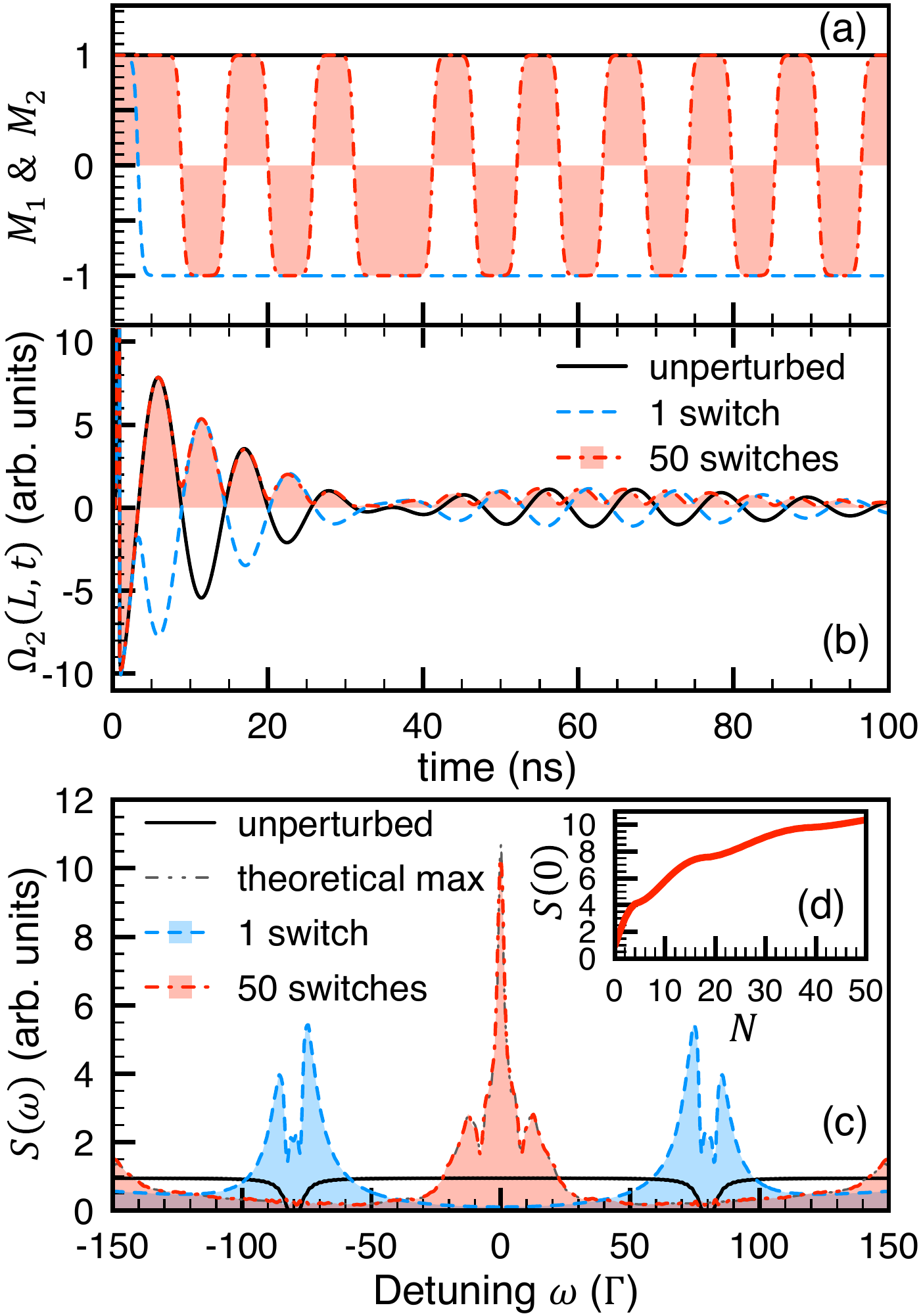}
\caption{\label{fig2}
(Color online) 
(a) The switching functions $M_1$ and $M_2$ for the effective single-target NFS $(M_1 = M_2)$.
(b) $\Omega_2 \left( L, t\right)$ with single (blue dashed line), multiple (red dashed-dotted-filled line), and without (black solid line) magnetic switching in the time domain. 
(c) The corresponding output x-ray frequency spectrum for $\xi_1=\xi_2=15$ and $\Delta=80\Gamma$. Gray dashed-dotted-dotted line depicts the theoretic maximum  spectrum.
(d) number of switching $N$-dependent $S\left( 0\right) $.
}
\end{figure}
A two-target system is reduced to an effective one-target system when  $\vec{B}_1$ and $\vec{B}_2$   are simultaneously switched, i.e., $M_1\left( t\right)  = M_2\left( t\right) $. 
Figure~\ref{fig2}  demonstrates the comparison between 
three kinds of perturbation for  $\Delta/\Gamma > \xi_j$. 
The switching functions $M_1\left( t\right) = M_2\left( t\right)$ and  $E_1\left( t\right) = \Omega_2 \left( L, t\right)$ are illustrated in Fig.~\ref{fig2}(a) and (b), respectively. Blue dashed, red dashed-dotted-filled, and black solid lines depict the field for single, multiple, and without magnetic switching, respectively. 
In the fifty-switch case the magnetic fields are inverted at each temporal node except the first.
In contrast to the  unperturbed NFS, the perturbed nuclear dynamics experiences time reversal due to the magnetic inversion at chosen instants when the x-ray intensity node occurs. This action effectively introduces a phase shift of $\pi$ in the scattered x rays.
Fig.~\ref{fig2}(c) depicts the corresponding frequency spectrum $S(\omega)$ for above cases.  
Remarkably, the phase flip in $\Omega_2 \left( L, t\right)$  turns the absorption dip into the enhanced spectral peak. With $\xi_1=\xi_2=15$ and $\Delta=80\Gamma$, the maximum  $S(\omega)=5.8$ for the single switching. 
$S(\omega)$ manifests both QB and DB. The spacing between two absorption dips in the unperturbed spectrum is $2\Delta$ associated with QB, and the spacing of the split around $\omega=\pm \Delta$ in the singly perturbed $S(\omega)$ is caused by DB. 
Moreover, fifty magnetic inversions turn alternating $\Omega_2 \left( L, t\right)$ into pulsating pattern, and so the frequency spectrum reveals both direct-current at  $\omega = 0$  and frequency doubling at $\omega = 2\Delta$ \cite{Heeg2017}. The full width at half maximum (FWHM) of the central peak of the fifty-switch $S(\omega)$ is $4\Gamma$. No significant deviation is observed for $d\leq5$ns.
When $N$ gets close to 50, $S(\omega)$ approaches the theoretically optimized  spectrum (gray dashed-dotted-dotted line) whose $\Omega_2\left( L, t >3.28 \mathrm{ns}\right) $ is completely flipped.
Fig. \ref{fig2}(d) shows that $S(0)$ approaches the theoretic maximum of $11$ when increasing the number of switchings $N \approx 50$ in 300ns. To achieve such high repetition rate, the magnetic switching \cite{Shvydko1996, Palffy2009, Liao2012a, Ramien2018} is potentially more preferable than reported mechanical solutions using piezo transducers \cite{Olga2014, Heeg2017}. The required repetition rate can be reduced by having smaller $\Delta$ but with the expense of lowering $S(0)$ due to the increase of magnetically unperturbable fraction before the first dynamical minimum.

\begin{figure}[b]
\includegraphics[width=0.43\textwidth]{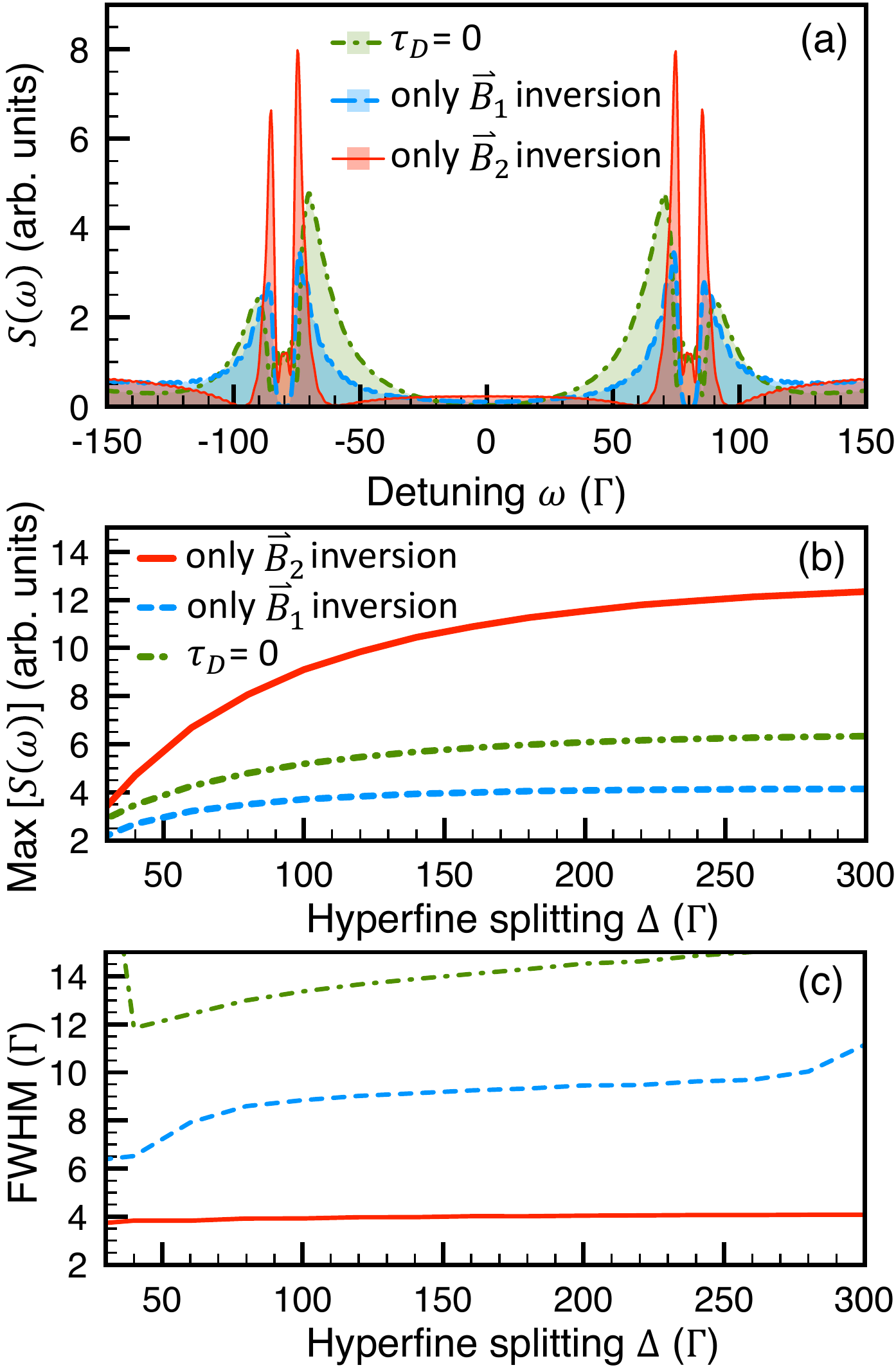}
\caption{\label{fig3}
(Color online) 
NFS frequency spectrum for $\Delta/\Gamma > \xi_j$.
(a) green dashed-dotted-filled line depicts the case of  simultaneous magnetic inversion ($\tau_D=0$) at $t=3.12$ns, and the red solid-filled (blue dashed-filled) line demonstrates the case where only $\vec{B_2}$ ($\vec{B_1}$) is inverted at $t=3.12$ns  with $\xi_1=\xi_2=30$.
(b) hyperfine splitting $\Delta$-dependent maximum $S(\omega)$. (c) $\Delta$-dependent FWHM.
}
\end{figure}
\begin{figure}[b]
\includegraphics[width=0.43\textwidth]{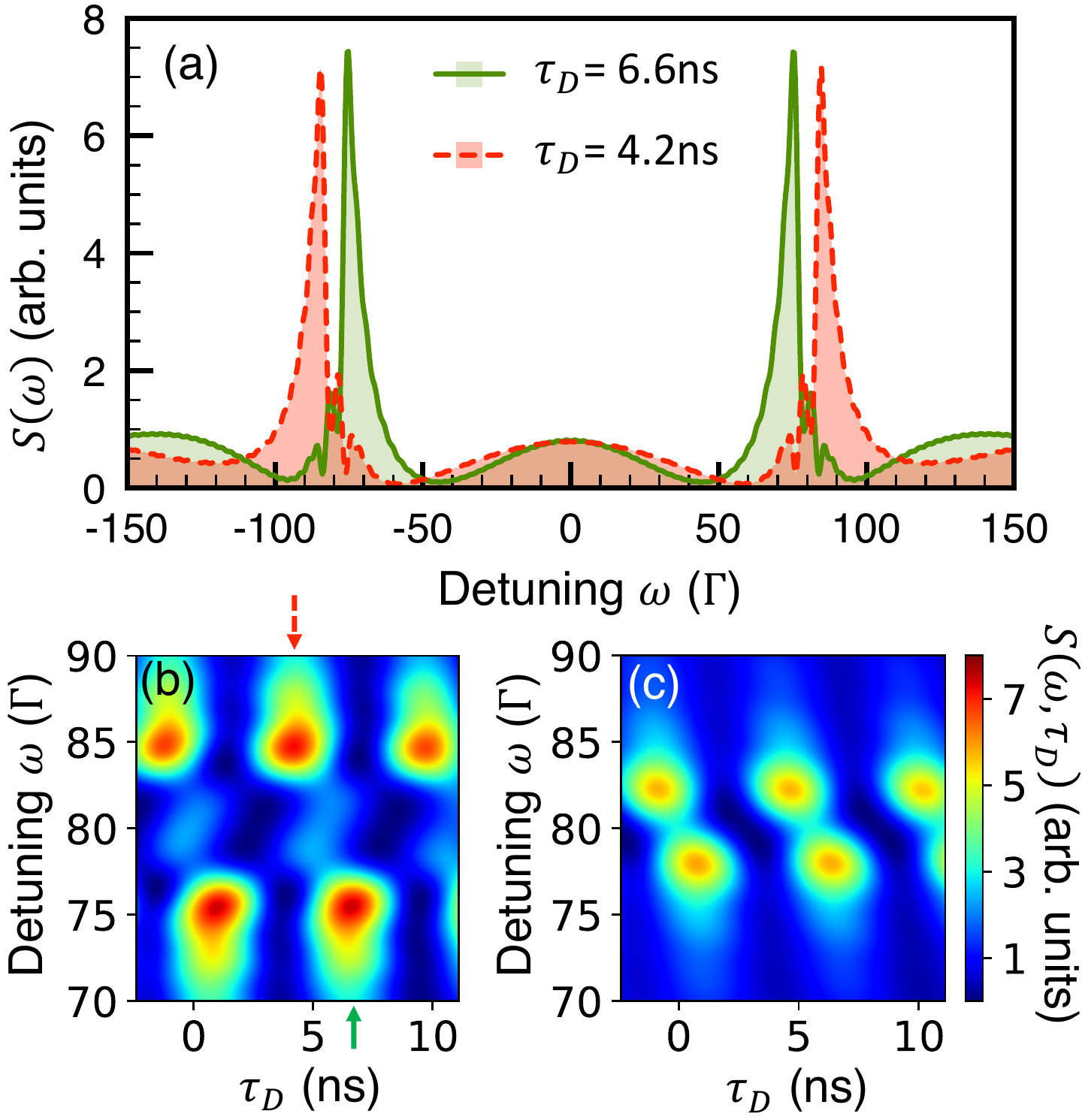}
\caption{\label{fig4}
(Color online) 
Perturbed NFS frequency spectrum for $\Delta/\Gamma > \xi_j$.
(a) red dashed-filled line (green solid-filled line) is the frequency spectrum for $\xi_1=\xi_2=15$, $t_1=3.28$ns and  $t_2=t_1+\tau_D$ where $\tau_D=4.2$ns (6.6ns)  indicated by the red dashed-downward (green upward) arrow in (b).
Two dimensional spectrogram for (b) $\xi_1=\xi_2=15$, and (c) $\xi_1=\xi_2=5$.  
(b) and (c) share the same color bar, and $\Delta=80\Gamma$ for all figures.
}
\end{figure}
Given $E_2(t)$ results from multi-path  interference, a two-target system renders  even more flexible controls possible. We first demonstrate  results for $\Delta/\Gamma > \xi_j$ in Fig.~\ref{fig3} and Fig.~\ref{fig4}. 
Fig.~\ref{fig3}(a) shows the perturbed $S\left( \omega\right) $  for three types of magnetic switching at $t=3.12$ns in a two-target system: 
(1) two magnetic fields are simultaneously inverted (green dashed-dotted-filled line), 
(2) only $\vec{B}_1$ inversion (blue dashed-filled line), and 
(3) only $\vec{B}_2$ inversion (red solid-filled line).
One can observe that each kind of switching leads to very different spectrum. In contrast to type (1) and (3), the overall spectral intensity of the  type (2) is lower, and there are dips  touching to zero around $\omega = \pm 80\Gamma$.
This is because in type (2) the upstream perturbation in the  first target generates a similar $S(\omega)$ to the blue dashed-filled line in Fig.~\ref{fig2}(b), but the downstream absorption by the second target reduces the resonant x-ray intensity.
As demonstrated by the red solid-filled line in Fig.~\ref{fig3}(a), the third kind of switching   not only  raises the spectral intensity to 8 times the input maximum, but also significantly narrows FWHM of each enhanced peak.
The FWHM of  spectral lines at $\omega = \pm 75\Gamma$ and $\omega = \pm 85\Gamma$ is about 3.4$\Gamma$ and 3.9$\Gamma$, respectively.
We illustrate the $\Delta$-dependent peak value of the perturbed $S(\omega)$  and FWHM for three types of switching in Fig.~\ref{fig3}(b) and Fig.~\ref{fig3}(c), respectively.  
Two quantities both show that the task of spectral narrowing can be further achieved by inverting only $\vec{B}_2$.
The maximum $S(\omega)$ of type (3) saturates with 13, and its FWHM remains insensitive to $\Delta$ and narrower than 4$\Gamma$.
In light of the perturbation  performed at the first temporal node of QB, 
shortening the QB period by increasing $\Delta$ reduces the
unperturbed part of NFS signal in time domain and therefore raises the spectral intensity in all types of magnetic switching.
Moreover, the difference between types (2) and (3) manifests a noncommutative effect  of magnetic perturbation in a two-target system. Since the exchange of perturbation ordering does not affect the interference between $W_1$ and $W_2$ in Eq.~(\ref{eq6}), the noncommutative effect is the consequence of the convolution term.
Inspired by the noncommutative property of our two-target system, we investigate the type (4) perturbation, namely, two magnetic fields are separately inverted with $\tau_D\neq 0$.
Based on the coherent nature of Eq.~(\ref{eq6}), a fine tuning of $\tau_D$ will lead to an alternation of spectral intensity at various resonant frequencies, which is sort of interference fringe.
Fig.~\ref{fig4} depicts the ebb and flow of $S\left( \omega\right) $ as a function of $\tau_D$ for $\xi_1=\xi_2=15$, $\Delta=80\Gamma$, switching $\vec{B}_1$ at $t_1=3.28$ns, and inverting $\vec{B}_2$ at $t_2=t_1+\tau_D$. The perturbed $S(\omega)$ for $\tau_D=4.2$ns ($\tau_D=6.6$ns) is depicted by red dashed-filled line (green solid-filled line) in Fig.~\ref{fig4}(a). Compared with the $S(\omega)$ of the synchronized switching in Fig.~\ref{fig2}(b), one can transfer a four-line spectrum to a double-line $S(\omega)$ by the type (4) switching. Moreover, we can control the relative height of the spectral line at $\omega = \pm 75\Gamma$ and $\omega = \pm 85\Gamma$. For $\tau_D=4.2$ns x-ray photons  concentrate around $\omega = \pm 85\Gamma$ with linewidth of $6\Gamma$. A later switching of $\vec{B}_2$ with $\tau_D=6.6$ns  inwardly redistributes NFS photons to $\omega = \pm 75\Gamma$. 
Fig.~\ref{fig4}(b) demonstrates the two dimensional $S(\tau_D, \omega)$ for $\xi_1=\xi_2=15$. $S(\tau_D, \omega)$ is a mirror image about $\omega=0$, and so we only show the result in $90\Gamma\geq \omega\geq 70\Gamma$ for the sake of a better visualization. We observe that $S(\tau_D, \omega)$ is gradually and periodically modulated along the variation of $\tau_D$. The red dashed-downward arrow and the green-upward arrow indicate the time delay for two cases in Fig.~\ref{fig4}(a).
In view that the tunable frequency components are determined by DB, Fig.~\ref{fig4}(c) depicts the change of using lower resonant thickness $\xi_1=\xi_2=5$ to relocate  two controllable spectral lines  at $\omega=77.5\Gamma$ and $82.5\Gamma$.

%
\begin{figure}[b]
\includegraphics[width=0.43\textwidth]{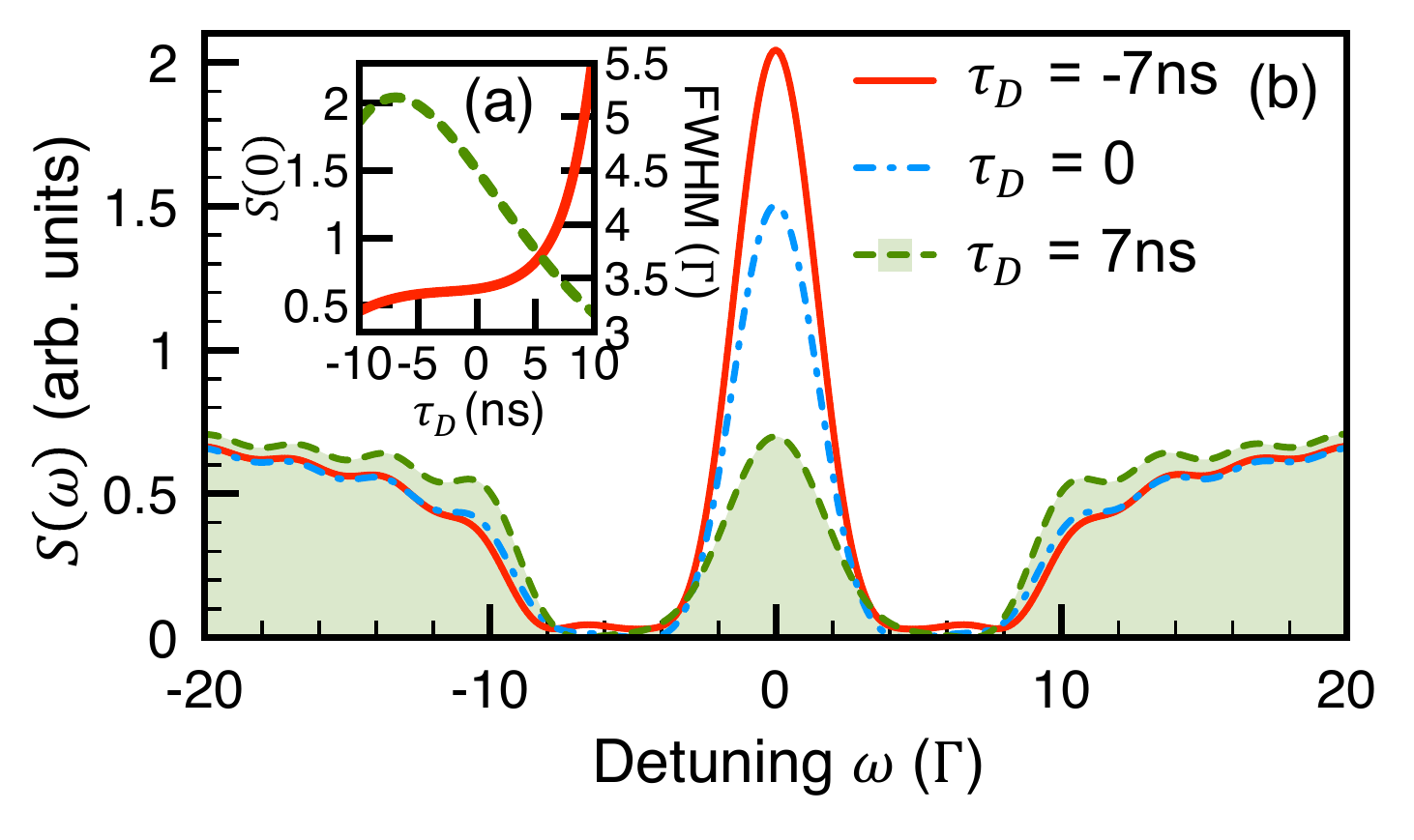}
\caption{\label{fig5}
(Color online)
(a) $\tau_D$-dependent $S\left( 0\right) $ (green dashed line) and FWHM of the single line spectrum (red solid line) for $\Delta/\Gamma < \xi_j$, where $\xi_1=\xi_2=15$, $\Delta=5\Gamma$ and $t_1=16.4$ns. 
(b) the perturbed NFS frequency spectrum.
Red solid, blue dashed-dotted and green dashed-filled lines are for $\tau_D=-7$ns, 0ns, and 7ns, respectively.}
\end{figure}
We now turn to discuss the region of $\Delta/\Gamma < \xi_j$ with type (1) and (4) switching. 
In Fig.~\ref{fig5}(a)  green dashed line (red solid line) illustrates $\tau_D$-dependent $S\left( 0\right) $ (FWHM) for $\xi_1=\xi_2=15$, $\Delta=5\Gamma$ and inverting $\vec{B}_1$ at $t_1=16.4$ns. The modulation of intensity is not periodic because DB dominates over QB in this region. Fig.~\ref{fig5}(b) demonstrates three cases for $\tau_D = -7$ns (red solid line), $\tau_D = 0$ (blue dashed-dotted line), and $\tau_D = 7$ns (green dashed-filled line). Each $S(\omega)$ turns into a single-line spectrum whose central peak is sandwiched by two wide dips touching the bottom. The central line is getting high and narrow when inverting $\vec{B}_2$ earlier than $\vec{B}_1$, namely, $\tau_D<0$. We observe $S(0)>2$ and FWHM $\approx 3.4\Gamma$ when $\tau_D = -7$ns. This demonstrates that a single enhanced resonant spectral line approaching the nature linewidth can be produced by the time-delayed control.

In conclusion we have demonstrated  versatile  controls over the  x-ray frequency spectrum in a magnetically perturbed  NFS system.  
The multiple magnetic switching in one-target NFS leads to a very narrow spectral line with tenfold intensity enhancement and FWHM about $4 \Gamma$.
A single switching in only the downstream target results in eight times magnification of spectral intensity and FWHM $< 4 \Gamma$ . This emphasizes the usefulness of our two-target system which has a  similar enhancement factor but avoids the need of high repetition rate of magnetic switching.
Moreover,  our time-delayed scheme allows for a flexible selection of enhanced spectral lines.
A combination of our scheme and a downstream nuclear resonant monochromator used in the synchrotron M\"ossbauer source \cite{Chumakov1990, Smirnov1997, Mitsui2009, Potapkin2012} will potentially lead to a  brighter  single-line x-ray source. This is generally useful for any flux-hungry precision spectroscopy using nuclear resonance, e.g., studies under extremely high pressure in a diamond anvil cell which is of great interest in geophysics \cite{Smirnov1997, Sturhahn2004, Mitsui2009, Potapkin2012, Potapkin2013}.

We thank A. P\'alffy, Y.-J. Chen, and  D.-A. Luh for fruitful discussions and suggestions.
This work is supported by the Ministry of Science and Technology, Taiwan (Grant No. MOST 107-2112-M-008-007-MY3 \& MOST 109-2639-M-007-002-ASP). 
W.-T.~L. is also supported by the National Center for Theoretical Sciences, Taiwan.

%
\bibliographystyle{apsrev}
\bibliography{NFSBS2A}

\end{document}